\documentclass[pre,preprint,double-spaced,aps,apssymb,showkeys,showpacs]{revtex4}

\usepackage{amsmath,amssymb}
\usepackage{graphicx}
\usepackage{dcolumn}
\usepackage{bm}
\usepackage{subfigure}
\usepackage{color}
\newcommand{\ber}{\begin{eqnarray}}
\newcommand{\eer}{\end{eqnarray}}
\newcommand{\bea}{\begin{equation}}
\newcommand{\eea}{\end{equation}}

\begin{document}

\title{Equilibrium stochastic dynamics of a Brownian particle in inhomogeneous space: derivation of an
alternative model}

\author{A. Bhattacharyay}
\email{a.bhattacharyay@iiserpune.ac.in}
\affiliation{Indian Institute of Science Education and Research, Pune, India}

\begin{abstract}
An alternative equilibrium stochastic dynamics for a Brownian particle in inhomogeneous space is derived. Such a dynamics can model the motion of a complex molecule in its conformation space when in equilibrium with a uniform heat bath. The derivation is done by a simple generalization of the formulation due to Zwanzig for a Brownian particle in homogeneous heat bath. We show that if the system couples to different number of bath degrees of freedom at different conformations then the alternative model is derived. We discuss results of an experiment by Faucheux and Libchaber which probably has indicated possible limitation of the Boltzmann distribution as equilibrium distribution of a Brownian particle in inhomogeneous space and propose experimental verification of the present theory using similar methods.
\end{abstract}
\maketitle
\par
Complex molecules like proteins, colloids etc in contact with a homogeneous heat bath form
a unique class of statistical systems. The most important characteristic feature of such
systems is that such systems
can be modelled by a Brownian particle (BP) with coordinate dependent damping \cite{lau,farago1,farago2}. In different conformations (structural configuration of a complex molecule e.g. a protein), the accessible surface area of such a molecule to bath degrees of freedom can be different. Coupling of the molecule to the uniform heat bath can then be conformation dependent \cite{expt1,expt2,fauc}. The motion of the complex molecule in its conformation space under equilibrium fluctuations is then similar to the motion of a BP in a multidimensional space with coordinate dependent damping. The source of inhomogeneity being damping, even in the absence of any homogeneity breaking potential (conservative force), the conformation space remains inhomogeneous. This is a situation for equilibrium statistics where there exists an inhomogeneous space in a homogeneous heat bath and no extra energetics is involved in maintaining the inhomogeneity of space. This is different from an inhomogeneous heat bath where to maintain the inhomogeneity of the heat bath some external agent (presence of at least a third system) must be involved and one cannot talk of equilibrium without taking into account the presence of that external agent and the entropy produced by that involvement.
\par
Relevant questions one can ask, given such a system, are - a. If the equilibrium probability distribution is the Boltzmann distribution when the damping is constant over space, what is the general probability distribution when damping is a function of space? This is an important question because the damping term cannot be accommodated in a Hamiltonian. b. When damping is inhomogeneous, what is the diffusivity? This is a relevant point because, damping being local can be inhomogeneous, but, diffusivity is a non-local quantity and should, therefore, be some average over space and time \cite{tupper}. c. How should one generalize the expression of stochastic force to decouple the local damping from the non-local diffusivity in equilibrium? As the present paper will elaborate, in the answer of this third question lie the answers of the previous two questions.
\par
In the conventional approach, coordinate dependent damping makes the equilibrium stochastic problem involve multiplicative noise. In so far existing literature on such theories the condition for equilibrium is considered to be the appearance of the Boltzmann distribution. This is questionable because the distribution does not involve damping which also breaks the homogeneity of space. Moreover, these standard procedures give rise to a host of other issues involving the dilemma of It\^o vs Stratonovich conventions \cite{ito,strat,san,sancho} etc.

\par
Let us first introduce the alternative approach here that has been introduced and elaborated in ref.\cite{ari1,ari2}. Consider the Brownian dynamics over a homogeneous space characterized by a damping constant $\Gamma$
\bea\nonumber
\frac{d x}{d t} = \frac{F(x)}{\Gamma} +\frac{g}{\Gamma}\eta(t).
\eea
In the above expression $F(x)=-\frac{\partial V(x)}{\partial x}$ is the conservative force and $g=\sqrt{2\Gamma k_BT}$ is the stochastic noise strength where $T$ is the temperature and $k_B$ is the Boltzmann constant. The Gaussian white noise $\eta(t)$ of unit strength has correlations $<\eta_i(t)>=0$ and $<\eta_i(t_i)\eta_j(t_j)>=\delta_{ij}\delta(t_i-t_j)$ as usual. The equilibrium probability distribution for the position of the BP given as $P(x)=Ne^{-V(x)/k_BT}$ where $N$ is a normalization constant. Taking an average on the above equation one gets $<\frac{dx}{dt}>=0$ where the noise term vanishes on average and the conservative force induced current identically vanishes on average using the Boltzmann distribution.
\par
Now consider a generalization of the above model where $\Gamma(x)$ and $g(x)$ are space dependent functions as
\bea\nonumber
\frac{d x}{d t} = \frac{F(x)}{\Gamma(x)} +\frac{g(x)}{\Gamma(x)}\eta(t).
\eea
Choosing $g(x)=\sqrt{2\Gamma(x)k_BT}$ and $\eta(t)$ being the same zero average white noise one conventionally obtains exactly the same Boltzmann distribution for the so called equilibrium of the particle. Take an average of the above equation again and $<F(x)/\Gamma(x)>\neq 0$ is the case now. The average over the noise term is then supposed to cancel out this non-zero current contribution in equilibrium.
\par
The noise term is locally Gaussian with a locally fixed width and zero average. A system to equilibrate over an inhomogeneous space like the given one the space must be of finite extent even in the absence of the conservative force. Given a long enough trajectory which revisits a particular coordinate many times, the noise term in principle can locally everywhere sample enough realizations to become increasingly small on average everywhere as the length of the trajectory increases. This simple fact means that given a locally Gaussian noise everywhere the noise term can tend towards zero on average everywhere over increasingly larger sampling. If the noise is locally Gaussian, this should be true independent of whatever technicality is involved in the multiplicative noise integration. Therefore, there would remain a residual current in the form of $<F(x)/\Gamma(x)>\neq 0$ as a result of having Boltzmann distribution and some {\it ad hoc} cancellation of this current in this so called equilibrium will be required.
  
\par In two previous papers \cite{ari1,ari2} we have discussed these issues and it has been shown on the basis of the sole consideration of the non-existence of any average current over inhomogeneous space in equilibrium that the stochastic force strength for equilibrium dynamics should be generalized to $g(x)=\Gamma(x)\sqrt{2k_BT/<\Gamma(x)>}$. This choice also explicitly decouples the local damping $\Gamma(x)$ from the global diffusivity $D={k_BT/<\Gamma(x)>}$ in the dynamics.

\par
With the above mentioned modified stochastic force strength which is a linear function of $\Gamma(x)$ the over-damped dynamics (Langevin dynamics) becomes a stochastic problem with additive noise and the distribution one gets readily as 
\bea
P(x) = N \exp{\left (\frac{<\Gamma(x)>}{k_BT}\int{dx \frac{F(x)}{\Gamma(x)}}\right )},
\eea 
where N is normalization constant. It is important to note that, according to this present model, the local diffusivity does not characterize equilibrium. The local diffusivity can always be defined in relation with the local damping over a suitable local average, but, that does not necessarily mean that it has to determine the equilibrium of the system. It is the global diffusivity of the system defined over the whole finite inhomogeneous space of the system that features in the equilibrium dynamics. Now, using this distribution it's straight forward to see that there is no average current in equilibrium and the distribution becomes the standard Boltzmann distribution when $\Gamma(x)$ is a constant. There are previous attempts to generalize Boltzmann distribution based on the entropy proposed by Tsallis and in this regard the ref.\cite{plst,tsal} are interesting works involving nonlinear Fokker-Planck equation.
\par
In this paper we are going to show a full derivation of the generalized Langevin dynamics including the inertial term and the above mentioned modified noise strength starting from a Hamiltonian that includes bath degrees of freedom as well as the system. We will follow here a formalism due to Zwanzig \cite{zwan}. It will be shown that if there exists a local conformation dependent coupling of the system to the harmonic bath the mesoscopic dynamics of the system will have the stochastic noise of the modified form. In the following we first summarize the results of the generalized Langevin dynamics (which has been obtained in details in ref.\cite{ari2}) in order to gain an idea about exactly what type of an equilibrium scenario is expected. After that, we would present a derivation of this generalized Langevin dynamics using the microscopic Hamiltonian. In the end we present a discussion.
\section{The generalized Langevin dynamics}
Consider the generalized Langevin dynamics including the inertial term as
\bea
m\frac{d^2x}{dt^2}=-m\Gamma(x)\frac{d x}{dt} + F(x) +\Gamma(x)\sqrt{\frac{2mk_BT}{<\Gamma(x)>}}\eta(t),
\eea
where m is the mass of the particle under consideration.
In ref.\cite{ari2}, the Fokker-Planck equation of the above model has been derived and the stationary distribution representing the equilibrium of the system is found out to be
\bea
P(x,v) = N\sqrt{\frac{m<\Gamma(x)>}{2\pi k_BT}}\frac{1}{\Gamma^{3/2}(x)}\exp{\left ( \frac{<\Gamma(x)>}{k_BT}\int_\infty^x{dx^\prime\frac{F(x^\prime)}{\Gamma(x^\prime)}}\right )}\exp{\left ( -\frac{mv^2<\Gamma(x)>}{2\Gamma(x)k_BT}\right )},
\eea
where $v=\frac{dx}{dt}$ and $N$ is an overall normalization constant.
The important thing to note here is that the velocity distribution is locally Gaussian with a position dependent width, which means that corresponding to this stationary distribution there exists a local temperature $T_{local} = T\Gamma(x)/<\Gamma(x)>$ which becomes the global temperature $T$ on an average over the whole inhomogeneous space. For example, if the inhomogeneous space is the internal space of a protein molecule, its different conformations would see a conformation dependent temperature but, as a whole, the temperature of the molecule over macroscopic time scales is that of the bath which is $T$. The molecule would remain in equilibrium with the bath as it samples through its conformation space being driven by the equilibrium fluctuations which indicates the $<\Gamma(x)>$ to be actually a weighted average which we would see to be indeed true as we look at equipartition of energy.
\par
Interesting to note that, although the conformations are characterized by local temperatures the average velocity at each conformation is zero because the velocity distribution is everywhere explicitly locally Gaussian with a zero mean. In the derivation of the probability distribution in ref.\cite{ari2} the standard procedure of setting the damping and the stochastic part dependent probability current to zero has actually been followed. Going by Einstein's definition of diffusivity, the local diffusivity in equilibrium of such a system is $D(x) = (\frac{k_BT\Gamma(x)}{<\Gamma(x)>})(\frac{1}{\Gamma(x)})=\frac{k_BT}{<\Gamma(x)>}$, which is identical to the global diffusivity. This is the most important signature of the present alternative theory over the existing ones. In the present context, equilibrium is characterised by this diffusivity and that entirely determines the stochastic nature of the problem. So, in equilibrium over an inhomogeneous space induced by coordinate dependent damping, it is not the temperature at all conformations which is constant, rather it is the diffusivity which remains constant. This observation, if is experimentally proven, can potentially have a much bigger role to play in the interaction of complex molecules involving barrier overcoming processes in equilibrium.
\par
Following ref.\cite{ari2}, let us have a look at the equipartition of energy and local power balance of the system. The local power balance in the presence of a conformation dependent temperature is indispensable in equilibrium. In the presence of the local power balance, there would remain no need for existence of any current to make unbalanced energy to flow between conformations (states) to maintain global energy conservation. In other words, the local power balance would ensure a local conservation of energy on noise average at each conformation in equilibrium.
\par
The equipartition can be easily seen to hold on average over the whole inhomogeneous space as
\ber\nonumber
<E_{KE}> &=& \frac{mN}{2}\sqrt{\frac{m<\Gamma(x)>}{2\pi k_BT}}\int dx{\frac{1}{\Gamma^{3/2}(x)}\exp{\left ( \frac{<\Gamma(x)>}{k_BT}\int_\infty^x{dx^\prime\frac{F(x^\prime)}{\Gamma(x^\prime)}}\right )}}\times \\
&& \int {dv v^2\exp{\left ( -\frac{mv^2<\Gamma(x)>}{2\Gamma(x)k_BT}\right )}} = \frac{k_BT}{2<\Gamma(x)>}\int{dx\Gamma(x)P(x)}=\frac{k_BT}{2},
\eer
where $P(x)$ is the distribution one obtains by locally integrating over all velocities. Consistent with our previous assessment we indeed get to see that the system as a whole maintains a constant temperature $T$ at which it can equilibrate with a heat bath.
\par
Note that, the computation of the equipartition implicitly presents the largest time scale (Macroscopic time scale) associated with the system over which the temperature seen by the system is the temperature of the heat bath. This is the time scale over which the system sees the whole of its allowed conformation space (inhomogeneous space) being sampled by its equilibrium fluctuations. Thermodynamically temperature is a quantity associated with a time scale below which one would see fluctuations in it. Equilibrium does not mean one would see the same temperature at all time scales. Compared to the above mentioned macroscopic time scale, the mesoscopic time scale is the one over which the coordinate dependent damping, corresponding local temperature and the stochastic form of the noise are defined. This particular time scale is there for every stochastic model defined at mesoscopic scales. For a homogeneous stochastic systems this mesoscopic time scale captures the temperature of the bath because the space is same everywhere whereas it is the macroscopic time scale (depending upon the extent of the inhomogeneosu space) over which the temperature of the bath is captured for an inhomogeneous stochastic system as considered here.
\par
The local power balance can also be seen to hold by computing the noise averaged input power locally as
\bea
E_{in}=\left <\left (\Gamma(x)\sqrt{\frac{2mk_BT}{<\Gamma(x)>}}\eta(t)\right )v(x,t)\right >,
\eea
 where the $v(t)$ is given by the dynamics and its stochastic part would only contribute to $E_{in}$ on a noise average. Thus,
\ber\nonumber
 E_{in} &=& \left ( \frac{2k_BT\Gamma^2(x)}{<\Gamma(x)>} \right ) e^{-\Gamma(x)t}\int_0^t{dt^\prime e^{\Gamma(x)t^\prime}<\eta(t)\eta(t^\prime)>}\\\nonumber
&=&  \left ( \frac{2k_BT\Gamma^2(x)}{<\Gamma(x)>} \right ) e^{-\Gamma(x)t}\int_0^t{dt^\prime e^{\Gamma(x)t^\prime}\delta(t-t^\prime)}\\
&=& m\Gamma(x){\overline {v^2(x)}} = E_{diss},
\eer
where $\overline {v^2(x)}$ is the mean square velocity at $x$. This balance is clearly overruling any energy current between conformations over noise average because the energy would remain locally conserved on noise average at each state although the states are characterized by local temperatures in equilibrium. The local temperature is an associated stochastic feature of the existence of the local damping when it exists as is taken in the model.

\par
Now, the question naturally arises is - does such a model have any microscopic basis? If it does have some microscopic basis then it should be derivable from some microscopic Hamiltonian where the above mentioned assumptions should reflect. The most important of all assumptions underlying the existence of such a model with coordinate dependent damping is the assumption that the system interacts with different amounts of bath degrees of freedom at different conformations (coordinates). In the following derivation of the model from microscopic Hamiltonian we will be using this pivotal assumption of differential coupling of the system with the bath.

\section{Microscopic derivation of the model}
Consider the Hamiltonian of the full microscopic system as
\bea
H = \frac{{\cal P}^2}{2} +\Phi(X) + \sum_i{\left [\frac{\omega_i^2}{2}\left
(q_i-\frac{\gamma(X)X}{\omega_i^2}\right )^2 + \frac{p_i^2}{2}\right ]}.
\eea
We consider the coupling constant $\gamma(X)$ is a function of the mesoscopic system's coordinate $X$ and
$\gamma(X)$ is independent of the bath oscillator index $i$. The conformation of the system is given by momentum ${\cal P}$ and coordinate $X$.
\par
Under this Hamiltonian the bath space is spanned by the
bath degrees of freedom (harmonic oscillators) specified by coordinates $q_i$ and momenta
$p_i$. We want to find out the stochastic dynamics of the mesoscopic system which is
harmonically coupled to the bath by integrating out $q_i$ and $p_i$. The $\Phi(X)$ is a potential causing the presence of some
conservative force to the mesoscopic system, $\omega_i$ is the frequency of the $i$th bath
oscillator where the masses of the bath and system are considered unity without any loss
of generality.
\par
The equation of motion for the system turns out to be
\bea
\frac{\dot{\cal P}}{f(X)} + \frac{1}{f(X)}\frac{\partial \Phi(X)}{\partial X} = \sum_i{D_i(t)},
\eea
where $f(X)=\frac{\partial \gamma(X)X}{\partial X}$, $D_i(t)=q_i(t) -
\frac{\gamma[X(t)]X(t)}{\omega_i^2}$ and an over-dot implies total time derivative.
The equation of motion for the $i$th bath degree of freedom is
\bea
\ddot{q_i}+\omega_i^2 q_i = \gamma(X)X,
\eea
which can be solved to give
\ber\nonumber
D_i(t) = q_i(t) - \frac{\gamma[X(t)]X(t)}{\omega_i^2} &=&
(q_i(0)-\frac{\gamma[X(0)]X(0)}{\omega_i^2})\cos{\omega_i t} +
\frac{p_i(0)}{\omega_i}\sin{\omega_i t}\\ &-&
\int_0^t{\frac{\cos{\omega_i(t-s)}}{\omega_i^2}}f(X){\cal P}(s)ds.
\eer

\par
Now, by replacing $D_i(t)$ in the equation of motion of the system one gets the dynamics of the system as
\bea
\frac{\dot{\cal P}}{f(X)} + \frac{1}{f(X)}\frac{\partial \Phi(X)}{\partial X} =
-\int_0^t{k(t-s){\cal P}(s)ds} +\phi(t),
\eea
where
\bea
k(t-s) = f(X)\sum_i{\frac{\cos{\omega_i (t-s)}}{\omega_i^2}}
\eea
and
\bea
\phi(t) = \sum_i{(q_i(0)-\frac{\gamma[X(0)]X(0)}{\omega_i^2})\cos{\omega_i t} +
\frac{p_i(0)}{\omega_i}\sin{\omega_i t}}.
\eea
\par
Let us first compute the quantity $k(t-s)$ at $X$ as
\ber\nonumber
k(t-s)&=&\lim_{\omega_{max}\to \infty}\int_0^{\omega_{max}}{d\omega
\frac{g(\omega,X)f(X)}{\omega^2}\cos{\omega(t-s)}}\\\nonumber &=&\lim_{\omega_{max}\to
\infty}2C\int_0^{\omega_{max}}{d\omega \cos{\omega(t-s)}}\\\nonumber
&=&\lim_{\omega_{max}\to \infty}2C\frac{\sin{\omega_{max}(t-s)}}{t-s}=2C\delta (t-s).
\eer
In the above computation we have taken the density of the oscillators of frequency $\omega$ at $X$ to be $g(\omega,X)=\frac{2C\omega^2}{f(X)}$ which is consistent with the scenario that at different conformations the system interacts with a varied number of bath degrees of freedom to have a conformation dependent damping and temperature. This choice of $g(\omega,X)$ is a local generalization of the density of oscillators as should be taken to generalize Zwanzig's formulation for the homogeneous space (additive noise) to non-homogeneous space \cite{zwan,jkb}. For the additive noise (homogeneous space) case (i.e. $\gamma(X)=\gamma$ a constant), the density of oscillators would be taken as $g(\omega)=\frac{2C\omega^2}{\gamma}$ in order to get a white noise following Zwanzig's formulation. Note that when $\gamma(X)\equiv \gamma$, we have $f(X)\equiv \gamma$ which clearly demonstrates the correspondence.
\par
Following the approach of Zwanzig, let us consider Maxwell-Boltzmann (MB) distribution of the initial coordinates $D_i(0)$ and momenta $p_i(0)$ of the bath degrees of freedom. However, the bath degrees of freedom we are considering here are the ones interacting with the system which is at its initial ($t=0$) position and momentum. When the system is in equilibrium with a state dependent temperature $\Gamma[X(0)]T/<\Gamma[X(0)]>$, guessing that $\Gamma(X) \propto f(X)$ we take the temperature of the distribution of the bath degrees of freedom at $t=0$ (as seen by the system in equilibrium with the bath) to be $f[X(0)]T/<f[X(0)]>$. Here, essentially, what being considered is that the particular bath degrees of freedom interacting with the system at a given conformation over a mesoscopic time scale are the only ones seen by the mesoscopic system at that local temperature. So, an average over bath degrees of freedom at $t=0$ should involve only these degrees of freedom which are seen by the mesoscopic system. Therefore, the statistics of the bath as seen by the system at the initial instant in the standard way \cite{jkb} is specified by the correlations involving the local temperature is given by the correlations $<p_i(0)> = <D_i(0)> = 0$,
$<p_i(0)p_j(0)> = \frac{k_Bf[X(0)]T}{<f[X(0)]>}\delta_{ij}$,
$<D_i(0)D_j(0)\omega_i\omega_j>=\frac{k_Bf[X(0)]T}{<f[X(0)]>}\delta_{ij}$,
$<D_i(0)p_j(0)> = 0$.
Using these averages at the initial instant it can be shown that

\ber\nonumber
<\phi(t)\phi(s)> &=& \left <\sum_{i,j}{\left (D_i(0)\cos{\omega_i t}+\frac{p_i(0)}{\omega_i}\sin{\omega_it}\right)\left (D_j(0)\cos{\omega_j s}+\frac{p_j(0)}{\omega_j}\sin{\omega_js}\right)} \right >\\\nonumber
&=& \sum_{i,j}{\frac{k_Bf[X(0)]T}{<f[X(0)]>\omega_i\omega_j}\delta_{ij}(\cos{\omega_it}\cos{\omega_js}+\sin{\omega_it}\sin{\omega_js})}\\\nonumber
&=& \frac{k_BT}{<f{X(0)}>}\sum_i{\frac{f[X(0)]]}{\omega_i^2}\cos{\omega_i(t-s)}}\\
&=&\frac{(k_BT)(2C\delta(t-s))}{<f[X(0)]>}=\frac{k_BTk(t-s)}{<f(X)>}.
\eer
Using the expressions for the second moment of $\phi(t)$ and $k(t-s)$, one can finally write Eq.10 in the form
\bea
\dot{\cal P}(t) = -\frac{\partial \Phi(X)}{\partial X} -Cf(X){\cal P}(t)+Cf(X)\sqrt{\frac{2k_BT}{C<f(X)>}}\eta(t),
\eea
where $\eta(t)$ is a white noise of unit strength and the factor of two in the damping term is cancelled by the half appearing due to integration over half of the delta function $\delta(t-s)$. Eq.15 is the equilibrium stochastic dynamics of a Brownian particle in an inhomogeneous space (the same as Eq.2) when we identify $m\Gamma(X)=Cf(X)$. A multi-dimensional extension of the same should follow the same line of logic.

\section{Existing experiment}
There is an experiment done by Faucheux and Libchaber \cite{fauc} to find out the variation of diffusivity of different Brownian particles moving under gravity between two confining planes. This experiment was based on direct visualization of trajectories of particles. In this experiment the whole 3-dimensional trajectory of a BP is projected on a horizontal $x-y$ plane. The diffusivity thus obtained on the transverse plane is an averaged quantity along the vertical direction. This average transverse diffusivity on horizontal planes $D_H$ is then plotted against a parameter which is the average height of a particle of a given radius from the lower confining plate. Obviously, this parameter scale is equivalent to the scale of particle size and is an appropriate one to plot vertically averaged horizontal diffusivity. It was shown that the diffusivity of a particle varies by as much as 2/3 of its bulk value at the closest proximity of the lower confining plate under experimental conditions. Here, the proximity is in terms of the parameter mentioned above which is a reasonably good measure.
\par 
A nice thing about such a system is that the vertical ($z$-direction) viscosity $\eta(z)$ is exactly known and the transverse (x,y) components of the viscosities were also used upto the approximate order $(r/z)^5$ where $r$ is the radius of the BP. The derivation of the present theoretical results completely rely on the conformation dependent coupling of the system with the bath degrees of freedom and no hydrodynamics of the bath giving rise to the variation in damping. However, the knowledge of the coordinate dependence of damping in this experimental system makes it an attractive one to probe the present results. If hydrodynamic effects are not too strong, most probably, such a system can reveal experimental results to prove the present theory.
\par
This experimental result, shown as early as in 1994, in the opinion of the present author, has clear indication of limitation of the Boltzmann distribution in an inhomogeneous space. The experimentally obtained horizontal average diffusivity $D_{H}$ as plotted against the average height of a particle should be a diffusivity obtained by averaging over the vertical $z$-direction. Now, the analytic expression of the $z$ dependence of the planar diffusivity $D_{||}(z)$ being known, if Boltzmann distribution $P_B(z)$ works, then $D_{H}=\int_0^t{dz D_{||}(z)P_B(z)}$ where $t$ is the vertical gap between two confining plates. It was clearly mentioned in this paper that this quantity does not match the experimentally observed average diffusivity. Then, by some particular double averaging (one local and the other global) the diffusivity is computed in this work which fits the experimental data (refer to the original paper \cite{fauc} for details).
\par
In the opinion of the present author, if the $D_{||}(z)$ is correct and the vertical distribution function is the Boltzmann distribution ($P_B(z)$) there is no reason why the direct averaging should fail. The experimental diffusivity has been extracted by nicely averaging over many sections of equal temporal stretches of a trajectory projected on horizontal plane ($x-y$ plane). This projection on $x-y$ plane is where the direct average in the vertical direction gets counted. The whole trajectory considered in this experiment is large enough to be able to properly see the available space. The small-time part of the mean square displacement showing an almost straight line (Fig.3 of ref.\cite{fauc}) indicates a very good averaging has been done. Now, this average cannot be different from a direct average of the planar diffusivity if $D_{||}(z)$ is correct and the distribution is the Boltzmann distribution. This is a beautiful experiment, most probably, giving a clear indication of a breakdown of the Boltzmann distribution in an inhomogeneous space where the inhomogeneity is caused by the space dependent damping. 
\par
Let us try to understand why such an experiment can be done to directly probe the modified Boltzmann distribution as presented in this paper before we elaborate on a proposal of a similar experiment. In the present experiment, it is a single Brownian particle whose motion near a wall is tracked. There can be multiple reasons for the variation in the damping near a wall for such systems. One of the reasons could be the differential coupling to the bath degrees of freedom if there exists a gradient of bath degrees of freedom near the wall. Or, there can most probably exist other hydrodynamic reasons, sticking of the boundary layers of water to the lower glass plate etc. Whatever be the microscopic origin of this damping variation with average height near the wall, if the energy lost by the particle to damping is retained by the fluid (bath) without loss and that transfer is a reasonably quick and local process, then it is a good system to probe proposed results. By reasonably quick we mean here that the time scale of this transfer is much smaller than the mesoscopic time scale. Such a situation should follow the same modified statistics as shown in this paper.    
  
\section{proposal of experiment}
We propose here two experiments one of which can be done exactly in the same way as was done by Faucheux and Libchaber for a direct visualization of the trajectory. In such an experiment one can obtain the (stationary) probability distribution of the vertical coordinate $z$ of the BP. Then one can fit this distribution to a modified Boltzmann distribution as proposed in this paper. The modified Boltzmann distribution $P_{MB}(z) = N \exp{(-V(z)/D)}$ is structurally the same as Boltzmann distribution with a different potential $V(z)=\int_0^z{dz^\prime ([m-m_w]g/6\pi r\eta(z^\prime))}$ where $m$ is mass, $r$ is radius of the particle under consideration, $m_w$ is the mass of water of same volume as that of the particle and $\eta(z)$ is the coordinate dependent viscosity for the motion in $z$-direction. The normalization constant $N$ should be found out for the whole span of the available space in the $z$-direction between confining plates. While fitting this distribution to the experimentally obtained one, the diffusivity D in the vertical direction can directly be measured in the experiment. The average global damping coefficient $<\Gamma(z)>=<6\pi r \eta(z)>$ will come out to be $k_BT/D$ where $T$ is the actual (known) temperature of the bath.
\par
Note that, this diffusivity $D=k_BT/<\Gamma(z)>$ in the vertical direction is dependent on the size or mass of the particle or in other words is dependent on the average height of the particle which has been considered as a parameter in the experiment of Faucheux and Libchaber. A direct measurement of the vertical diffusivity $D$ of the system, for example, above and below the average height of the particle would be much more interesting result in such an experiment. If the present theory works here, one should get to see the vertical diffusivity remains the same above and below the average height of the particle although there is a variation in the damping. The important thing to keep in mind in the evaluation of this vertical diffusivity is that it is not the average diffusivity $<K_BT/6\pi r\eta(z)>$ in this direction, rather, it is a diffusivity determined by dividing the $k_BT$ by $<6\pi r\eta(z)>$. It is the diffusivity corresponding to the average viscosity $<\eta(z)>$ and is not the average of diffusivity in the vertical direction.  
\par
The transverse planer diffusivity that Faucheux and Libchaber have plotted against the average height, according to the present theory, should be averaged using this new distribution in the vertical direction. The viscosity on horizontal planes at a given height being uniform everywhere the conventional theory of Brownian motion would work on these planes. On the transverse plane the distribution is the Boltzmann distribution if any planar confinement exists through a conservative potential. Otherwise, it is just conventional open diffusion on the transverse planes with diffusivity $D_{||}(z)$. Now, the average horizontal diffusivity should be $D_{H}=\int_0^t{dz D_{||}(z)P_MB(z)}$. If the direct average in this way matches the experimentally obtained one then that will also be the proof of existence of the modified distribution.
\par
A second experiment of different type where the present theory should be more applicable would be the one involving a complex molecule e.g. a protein or a star polymer immersed in a fluid at a constant temperature and staying far away from the boundary. A direct tracking of the conformations of the system over time should be done here. Then, the accessible surface area of the system to bath degrees of freedom can be estimated out from those conformations. This accessible surface area to bath degrees of freedom can constitute the coordinates of the one-dimensional space over which the diffusion of the system can be defined. This diffusion over the coordinate space of accessible surface area would then be the one-dimensional diffusion under the equilibrium fluctuations.
\par
One can directly probe here the constancy of diffusivity over the whole conformation space although the damping can change from conformation to conformation. These are systems probably where the damping variation with conformation is not as clearly known as that in the experiment of Faucheux and Libchaber. However, if the diffusivity comes out to be a constant in such an experiment, then one can find out the effective potential from the fitting of the graph of probability distribution to the experimentally obtained one in this one-dimensional space. A comparison of the effective potential with the actual one can then give a conformation dependent viscosity back if that exists. The weighted average of this conformation dependent viscosity then should again give back the same constant diffusivity to complete the self-consistency check.

\section{Discussion}
In the present paper we have considered a mesoscopic system in contact with a heat bath where the coupling of the system with the heat bath depends on the conformation of the system. Starting from this microscopic scenario of conformation dependent coupling of a mesoscopic system with a heat bath we have derived the mesoscopic dynamics of the system which is in the structure we have expected. The form of the mesoscopic dynamics actually differs from the other conventionally taken Langevin dynamics where the stochastic force strength is assumed as $\sqrt{2m\Gamma(x)k_BT}$.
\par
We introduced the local MB distribution for the bath degrees of freedom just before Eq.14 and that is an assumption. It does not anyway mean that because of making that assumption the distribution of the mesoscopic system would also be a similar one. Rather the distribution of the mesoscopic system has to be found out once we know the dynamics of the system after integrating out the bath degrees of freedom. Since, the dynamics of the system comes out in the expected form corresponding to which we know the distribution we can write down the distribution of the whole system (comprising of the bath and the system) in a product form. Integrating out the bath part then (when its distribution is suitably normalized), we can always recover the distribution of the system from such a distribution of the whole system. Note the interesting limit where $\Gamma(x)$ becomes independent of $x$ the distribution for the system we actually get is standard MB for the system and the bath alike as is obvious over homogeneous space.
\par
The important point here to make is we cannot {\it a priori} assume the distribution to be standard MB for a system which has conformation dependent coupling to a bath. This has to be found out explicitly from the mesoscopic dynamics and the assumption of standard MB distribution in equilibrium would be a wrong one to make. It works for the homogeneous damping does not mean that it should be a general procedure. Note that, considering the system and the bath together does not mean that one is at thermodynamic limit. The system goes to thermodynamic limit when its boundary goes to infinity keeping the densities finite. But, that is not the case here and we are within a mesoscopic domain.
\par
The generalized Langevin dynamics with the conventional noise strength $\sqrt{2m\Gamma(x)k_BT}$ gives a standard MB distribution for supposed equilibrium and it has all the merits like equipartition and local energy balance as shown here for our modified model.The reason for this correspondence most probably lies in the fact that, the conventional generalized Langevin dynamics can also be converted to an additive noise problem by rescaling the velocity by $\sqrt{\Gamma(x)}$ which is done in ref.\cite{ari2} by $\Gamma(x)$. This indicates that to fix the correct stochastic noise strength one has to take into account the over-damped dynamics limit. Here, the crucial characteristic feature that the conventional method lacks is that one cannot directly go to the over-damped dynamics by dropping the inertial term from the dynamics which, on the contrary, is straightforwardly doable with our model. 
\par
In the conventional case one has to augment the dynamics that results from dropping the inertial term with other {\it ad hoc} terms to cancel current. There is absolutely no need for such things for our model. In fact, our model with the noise strength $\Gamma(X)\sqrt{\frac{2mk_BT}{<\Gamma(X)>}}$ where the noise is a linear function of the damping is the only form which would result in an additive noise dynamics for equilibrium in the over-damped regime. This over-damped dynamics is absolutely free of any spurious current. All other noise forms including the standard one would remain multiplicative noise dynamics in the over-damped regime and that generality most probably indicates the common non-equilibrium character of them.  
\par
We would like to conclude by mentioning that the knowledge of the exact equilibrium probability distribution of a class of systems is not only required for the systems in equilibrium, rather, is required for all the estimations in the linear response regime which includes a lot more phenomena. For example, if experimentally proved, the conformation dependent temperature of complex molecules under equilibrium fluctuations can influence their reactions which cannot be explained by the Boltzmann statistics. If a wrong distribution is used for equilibrium the same gets involved in the weakly non-equilibrium regime as well giving rise to anomalies which should have not appeared with a proper theory. One of the long standing unsolved issue is Levinthal's paradox in protein folding. Possibly, a modification of the equilibrium distribution of mesoscopic systems holds the clue as to how a path is cut over a rugged energy landscape to make the system find the global minimum corresponding to the native fold of a protein. 
\\

\end{document}